\documentclass[useAMS,usenatbib, letterpaper]{mn2e}
\pdfoutput=1 

\usepackage{graphicx}
\usepackage{amssymb}
\usepackage{natbib}
\usepackage{aas_macros}

\title[Physical Conditions of the Gas in an ALMA {[CII]}-identified Submillimetre Galaxy at $z = 4.44$]{Physical Conditions of the Gas in an ALMA {[CII]}-identified Submillimetre Galaxy at $z = 4.44$} 
\author[M. Huynh et al.] {M.T.~Huynh,$^1$\thanks{E-mail: minh.huynh@uwa.edu.au}
R.P. Norris,$^2$ K.E.K. Coppin,$^3$ B.H.C. Emonts,$^2$ R.J. Ivison,$^4$  \newauthor 
N. Seymour,$^2$ Ian Smail,$^5$ V. Smol{\v c}i{\'c},$^{6,7,8}$ A.M. Swinbank,$^5$ W.N. Brandt,$^9$ \newauthor 
S.C. Chapman,$^{10}$ H. Dannerbauer,$^{11}$ C. De Breuck,$^{6}$  T.R. Greve,$^{12}$ J.A. Hodge,$^{13}$  \newauthor
A. Karim,$^5$ K.K. Knudsen,$^{14}$ K.M. Menten,$^{15}$  P.P. van der Werf,$^{16}$ F. Walter$^{13}$\newauthor and A. Weiss$^{15}$ \\
$^{1}$ International Centre for Radio Astronomy Research, M468, University of Western Australia, Crawley, WA 6009, Australia \\
$^{2}$ CSIRO Astronomy and Space Science, PO Box 76, Epping, NSW, 1710, Australia \\
$^{3}$ Department of Physics, McGill University, 3600 Rue University, Montreal, QC H3A 2T8, Canada \\
$^{4}$ UK Astronomy Technology Centre, Royal Observatory, Blackford Hill, Edinburgh EH9 3HJ, UK \\
$^{5}$ Institute for Computational Cosmology, Department of Physics, Durham University, Durham DH1 3LE, UK \\
$^{6}$ European Southern Observatory, Karl-Schwarzschild-Stra{\ss}e 2, D-85748 Garching b. Muenchen, Germany \\ 
$^{7}$ Argelander Institut for Astronomy, Auf dem Hugel 71, Bonn D-53121, Germany \\
$^{8}$ Physics Department, University of Zagreb, Bijeni{\v c}ka cesta 32, 10002 Zagreb, Croatia \\
$^{9}$ Department of Astronomy and Astrophysics, Pennsylvania State University, 525 Davey Lab, University Park, PA 16802, USA \\
$^{10}$ Institute of Astronomy, University of Cambridge, Madingley Road, Cambridge, CB3 0HA, U.K. \\
$^{11}$ Universit{\"a}t Wien, Institut f{\"u}r Astrophysik, T{\"u}rkenschanzstra{\ss}e 17, 1180 Wien, Austria \\
$^{12}$ Department of Physics \& Astronomy, University College London, Gower Street, London WC1E 6BT, UK \\
$^{13}$ Max-Planck Institute for Astronomy, K{\"o}nigstuhl 17, D-69117 Heidelberg, Germany \\
$^{14}$ Department of Earth and Space Sciences, Chalmers University of Technology, Onsala Space Observatory, SE-43992 Onsala, Sweden \\
$^{15}$ Max-Planck-Institut f{\" u}r Radioastronomie, Aus dem H{\" u}gel 69, 53121 Bonn, Germany \\
$^{16}$ Leiden Observatory, Leiden University, PO Box 9513, 2300 RA Leiden, The Netherlands}

\begin{document}



\maketitle

\label{firstpage}

\begin{abstract}

We present $^{12}$CO(2--1) observations of the submillimetre galaxy ALESS65.1 performed with the Australia Telescope Compact Array at 42.3 GHz. 
A previous ALMA study of submillimetre galaxies in the Extended {\it Chandra} Deep Field South detected  [CII] 157.74 $\mu$m emission from this galaxy at a redshift of $z = 4.44$. 
No $^{12}$CO(2--1) emission was detected but we derive a firm upper limit to the cold gas mass in ALESS65.1 of $M_{\rm gas} < 1.7 \times 10^{10}$ ${\rm M}_\odot$. The estimated gas depletion timescale is $<50$ Myr, which is similar to other high redshift SMGs, and consistent with $z > 4$ SMGs being the likely progenitors of massive red-and-dead galaxies at $z > 2$. The ratio of the [CII], $^{12}$CO and far-infrared luminosities implies a strong far-ultraviolet field of $G_0 \gtrsim 10^3$, as seen in Galactic star forming regions or local ULIRGs. The observed $L_{\rm [CII]}/L_{\rm FIR} = 2.3 \times 10^{-3}$ is high compared to local ULIRGs and, combined with $L_{\rm [CII]}/L_{\rm CO} \gtrsim 2700$, it is consistent with ALESS65.1 either having an extended (several kpc) [CII] emitting region or lower than solar metallicity.

\end{abstract}


\section{Introduction}

Submillimetre galaxies (SMGs) are ultraluminous, dusty starbursting systems with extreme star formation rates (SFRs)  of 100 -- 1000 $\rm{M}_\odot$ yr$^{-1}$ (e.g. \citealp{blain2002}). These SMGs have typical redshifts of $z \sim 2$--3 (e.g. \citealp{chapman2005, wardlow2011,yun2012,smolcic2012}), but an increasing number of higher redshift SMGs are being found. 
The $z > 4$ SMGs represent candidates for the most intense star formation phase of the massive red galaxies seen at $z > 2$ \citep{cimatti2008, marchesini2010}.
This high redshift tail of the SMG distribution was long under-represented in SMG redshift surveys, mostly because they lie below the sensitivity limits of the 
radio interferometer surveys used to identify SMGs. 
However, in the last few years, over a dozen of these sources have been reported  \citep{capak2008,capak2011,daddi2009b,daddi2009a,coppin2009,knudsen2010, carilli2010, carilli2011,riechers2010,smolcic2011,cox2011,combes2012, walter2012}, and recent ALMA redshift surveys have doubled these numbers (Weiss et al. submitted; Vieira et al. submitted).
Detailed studies of the star formation and gas content in these $z > 4$ SMGs are still rare, but are needed to provide a unique insight into the earliest phases of the growth of massive elliptical galaxies.

The most accessible tracer of cold gas in galaxies is $^{12}$CO, which has been observed extensively in SMGs (e.g. \citealp{greve2005,tacconi2006,bothwell2013}). Detections of $^{12}$CO in $z > 4$ SMGs have shown they are gas rich systems with sufficient reservoirs ($M_{\rm gas} > 10^{10}$ $\rm{M}_\odot$) to sustain the extreme star formation rates of $\sim$ 1000 $\rm{M}_\odot$ yr$^{-1}$ for only short time scales (10s of Myrs) \citep{coppin2010, riechers2010}, unless the gas is replenished. This is consistent with high redshift SMGs being the progenitors of the luminous red galaxies seen at $z > 2$.

The $^2P_{3/2}$ -- $^2P_{1/2}$ fine structure line of singly ionised carbon at 157.74 $\mu$m (here-after [CII]) has emerged as a powerful alternative line for studying the ISM in high redshift sources.  It can represent up to 1\% of the bolometric luminosity of star forming galaxies (e.g. \citealp{crawford1985, stacey1991}). This line emission arises mainly from the photodissociation regions (PDRs) that form at the edges of molecular clouds illuminated by the UV photons of young-massive stars,  but a significant contribution can also come from HII regions and the more diffuse warm interstellar medium \citep{madden93,heiles94}. The [CII] line therefore provides an important probe of the gas content and star formation processes in a galaxy. 

A recent ALMA Cycle 0 study of 126 submillimetre sources located in the LABOCA Extended {\it Chandra} Deep Field South (LESS, \citealp{weiss2009}; Karim et al. 2012; Hodge et al. 2013 submitted) resulted in the serendipitous identification of [CII] line emission from two SMGs (\citealp{swinbank2012}, hereafter S12). The average [CII[/far-infrared luminosity ratio of these two SMGs is $\sim$ 0.0012 $\pm$ 0.0004, roughly ten times higher than that observed in local ultraluminous infrared galaxies, which has been interpreted as evidence that their gas reservoirs are more extended (S12). 
The large extent of SMGs is supported by other observations, such as extended radio morphologies (e.g. \citealp{chapman2004,biggs2008}), extended H-$\alpha$ morphologies (e.g. \citealp{swinbank2006}), large $^{12}$CO(1-0) sizes \citep{ivison2010a, ivison2011, thomson2012, hodge2012},  lack of silicate absorption in mid-infrared spectra \citep{menendez-delmestre2009}, and less reddened broad-band mid-infrared colours \citep{hainline2009}.   High [CII]/far-infrared ratios  (\citealp{stacey2010, ivison2010b}; S12) add to this mounting evidence that star-formation in SMGs takes place in a region larger than the compact nuclear starbursts of local ULIRGs. 

In this paper we present $^{12}$CO(2--1) observations of one of the ALMA detected SMGs, ALESS J033252.26-273526.3 (hereafter ALESS65.1). We adopt the standard $\Lambda$-CDM cosmological parameters of $\Omega_{\rm M} = 0.27$,  $\Omega_{\rm \Lambda} = 0.73$, and a Hubble constant of 71 km s$^{-1}$ Mpc$^{-1}$ throughout this paper.

\section{Observations and Results}

The $^{12}$CO(2--1) line ($\nu_{\rm rest}$ = 230.538 GHz) in ALESS65.1 (RA(J2000) = 03 32 52.26, Dec(J2000) = $-$27 35 26.3) (S12) was observed over a period of four consecutive nights, 9 -- 12 August 2012,  with the Australia Telescope Compact Array (ATCA), using the Compact Array Broadband Backend (CABB). The array was in the most compact five-antenna configuration, H75, which has a maximum baseline of 89m and two antennas set along a northern spur. The hybrid configuration allows good $(u,v)$ coverage to be obtained for integrations less than the full 12 hour synthesis.  We centered the 7mm IF1 receiver on 42.343 GHz, the expected frequency of the $^{12}$CO(2--1) line emission given the [CII] redshift of $z = 4.4445$ derived by S12. The 2GHz wide bandwidth of CABB results in a frequency coverage of 41.3 to 43.3 GHz, covering $^{12}$CO(2--1) emission between $z =$ 4.32 -- 4.58.  The weather was good to average, with rms atmospheric path length variations of 100 to 400 microns throughout the run, as measured on the 230m baseline ATCA Seeing Monitor \citep{middelberg2006}.
The system temperature was 140 to 250 K throughout the four nights. Weather and atmospheric conditions can induce temporal fluctuations across the wide CABB band, so, following \cite{emonts2011}, a bandpass calibration scan was acquired at the beginning, middle and end of each 8 hour night. Phase and amplitude calibration information was acquired with 2 minute scans on PKS 0346$-$279 every 10 minutes and pointing checks performed on the same source every hour. Flux density calibration was performed on Uranus at the beginning of the nights, when it was at an elevation of roughly 55 degrees. The uncertainty in the flux density calibration using the standard {\sc MIRIAD} model of Uranus is estimated to be 30\% \citep{emonts2011}. 

The data were calibrated, mapped and analysed using the standard {\sc MIRIAD} \citep{sault1999} and {\sc KARMA} \citep{gooch1996} packages.  The synthesized beam from natural weighting is 14.0 $\times$ 9.9 arcsec. A total of about 20 hours on-source integration time was obtained over the 4 $\times$ 8 hour nights. 
ALESS65.1 was not detected in the 42.3 GHz continuum map from the full CABB band, which achieves an rms noise level of 11 $\mu$Jy beam$^{-1}$. 

The resultant channel noise in the 1 MHz (7.1 km s$^{-1}$) wide spectrum is $\sim$ 0.43 mJy beam$^{-1}$, consistent with other comparable 7mm ATCA/CABB surveys (e.g. \citealp{coppin2010, emonts2011}). 
The visibilities were resampled to velocity resolutions of 50, 100 and 200 km s$^{-1}$ and each cube was examined for an emission line near the ALMA position. A $\sim$4$\sigma$ spike at the position of ALESS65.1 and offset from the [CII] emission by $\sim$ 280 km s$^{-1}$ was examined in detail. It was deemed a noise spike because it is only present in one channel in all three binned images and is present only in the last night of data when each night is imaged separately. We flag the spike (20 native channels) in the last night of data and recombined all four nights of data to derive the final spectra. The spectra at the source position in the 100 and 200 km s$^{-1}$ binned cubes (Figure 1) have an rms of  0.11 and 0.08 mJy beam$^{-1}$, respectively. Assuming a line FWHM identical to the [CII] emission line (470 km s$^{-1}$, S12) and adopting a 3$\sigma$ limit, the $^{12}$CO(2--1) line intensity of ALESS65.1 is $I_{\rm CO(2-1)}$ $< 0.11$ Jy km s$^{-1}$.

\begin{figure}
\centering
\includegraphics[width=0.99\columnwidth]{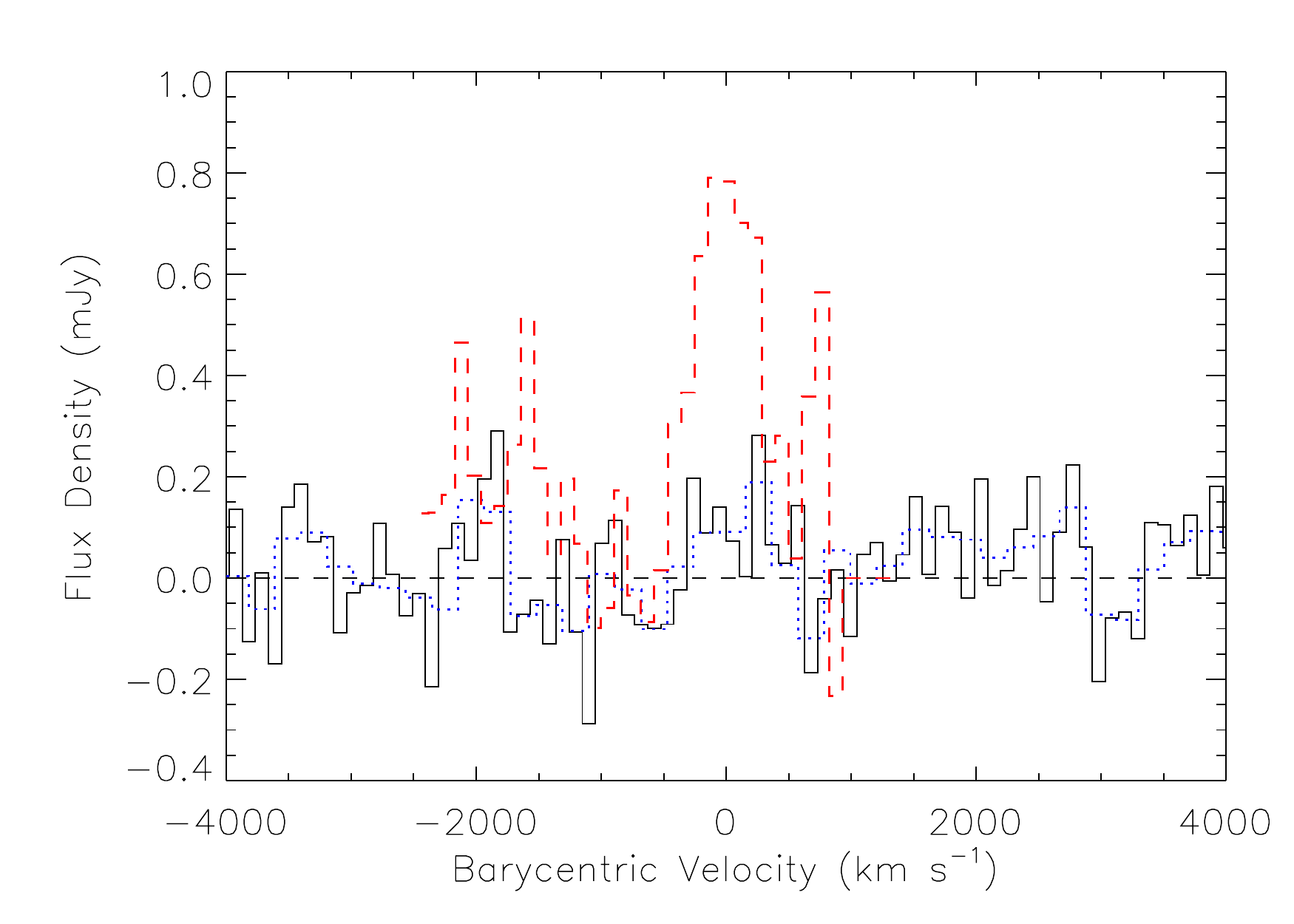}
\caption{$^{12}$CO(2-1) spectrum of ALESS65.1 binned into 100 km s$^{-1}$ channels and extracted at the ALMA position. The red dashed line shows the ALMA [CII] spectrum from S12, binned to a similar channel width (120 km s$^{-1}$) and with the flux density divided by 15 for clarity. The blue dotted line shows the spectrum binned into 200 km s$^{-1}$ for maximum sensitivity. We conclude that no line was detected to a 3$\sigma$ limit of 0.24 mJy beam$^{-1}$ per 200 km s$^{-1}$ channel.}
\label{fig:spectra}
\end{figure}

\section{Discussion}

The observed and derived properties of ALESS65.1 are summarised in Table 1. To estimate the cold molecular gas content in ALESS65.1 we calculate upper limits to the line luminosity and total cold gas (H$_2$ + He) mass from the CO(2--1) flux density limit. Following Solomon and Vanden Bout (2005), the line luminosity upper limit is $L'_{\rm CO(2-1)}$ $<$ 2.2 $\times$ $10^{10}$ K km s$^{-1}$ pc$^2$. If we assume the gas is thermalised (i.e. intrinsic brightness temperature and line luminosities are independent of $J$ transition), so $L'_{\rm CO(2-1)}$ = $L'_{\rm CO(1-0)}$, and a CO-to-H$_2$ conversion factor $\alpha = 0.8$ ${\rm M}_\odot$ (K km s$^{-1}$ pc$^2$)$^{-1}$, which is appropriate for ULIRGs (e.g. \citealp{downes1998}, but see \citealp{bothwell2012}), the upper limit on the total cold gas mass is $M_{\rm gas} < 1.7 \times 10^{10}$ ${\rm M}_\odot$. This is consistent with the gas mass found in other $z > 4$ SMGs \citep{schinnerer2008, daddi2009b, coppin2010, walter2012} and the typical [CII]/$M_{\rm gas}$ ratio at high redshift (S12). 

The total baryonic mass of ALESS65.1 can be derived by combining the gas and stellar mass estimates for the system. The stellar mass of the system was estimated from the rest-frame absolute H-band magnitude to be $M_* \sim 9 \times 10^{10}$ ${\rm M}_\odot$ (S12), so the gas fraction is modest with $M_{\rm gas}/M_* \lesssim 0.2$. The total baryonic mass $M_{\rm bary}$ = $M_{\rm gas} + M_*$ is $\sim$ 9 -- $11 \times 10^{10}$ ${\rm M}_\odot$. This is consistent with the dynamical mass for the system, based on the [CII] linewidths and spatial extent, of $M_{\rm dyn} \sin^2(i) \sim (3.4 \pm 1.8) \times 10^{10}$ ${\rm M}_\odot$ (S12).

ALESS65.1 is detected in the 870$\mu$m continuum by ALMA and weakly detected in the far-infrared (FIR) by {\it Herschel} (S12). A fit to the IR SED resulted in a restframe IR (8 -- 1000 $\mu$m)  luminosity $L_{\rm IR} = (2.0 \pm 0.4) \times 10^{12}$ ${\rm L}_\odot$ (S12), which corresponds to a star formation rate of $\sim$ 340 ${\rm M}_\odot$ yr$^{-1}$ using the conversion of \cite{kennicutt1998}. The gas depletion timescale $\tau = M_{\rm gas} /{\rm SFR} < 50$ Myr is similar to the gas depletion rates of other high redshift SMGs \citep{schinnerer2008,coppin2010}. This short timescale provides further evidence that $z > 4$ SMGs have the gas consumption timescales necessary to be the progenitors of red-and-dead ``ellipticals" at $z > 2$. 
 
Next we examine the physical conditions of the gas in ALESS65.1 using the [CII] detection and $^{12}$CO(2--1) limit. The  $L_{\rm [CII]}/L_{\rm FIR}$  versus $L_{\rm CO(1-0)}/L_{\rm FIR}$ diagram is a powerful diagnostic as these two ratios are sensitive to gas density $n$ and the incident far-ultraviolet (FUV) flux $G_0$ \citep{stacey1991}. Figure 2 shows ALESS65.1 compared with other low and high redshift galaxies, and solar metallically PDR model curves \citep{kaufman1999}. This diagram can be used to roughly estimate both $n$ and $G_0$ for a galaxy, but there are some caveats, as outlined in \cite{debreuck2011}. These are: (i) the $^{12}$CO luminosity is that of the ground state rotational line, (ii)  [CII] emission from the diffuse ionised medium and cosmic-ray-heated gas is assumed to be small compared to that from PDRs, and (iii) AGN and their related X-Ray Dissociation Regions (XDRs) are assumed to not contribute significantly to the FIR and [CII] luminosity. To be consistent with both \cite{debreuck2011} and \cite{stacey2010} in Figure 2 we assume $L_{\rm CO(2-1)} / L_{\rm CO(1-0)} = 7.2 $, which is 90\% of its value if the gas was fully thermalised and optically thick. Cosmic ray rates are greater in starbursts compared to normal galaxies but this does not seem to result in higher [CII]/CO ratios, so cosmic ray ionization does not appear to dominate the [CII]/CO ratio \citep{debreuck2011}, at least for local galaxies.  ALESS65.1 is not detected in the 250ks {\it Chandra} X-Ray observations of this region \citep{lehmer2005}, so it is not a luminous QSO ($L_{\rm 3-44 keV} \lesssim 2$--3$ \times 10^{44}$ erg s$^{-1}$, for $N_{\rm H} = 0$ -- 10$^{23.5}$ cm$^{-2}$). The maximum likely AGN contribution to the FIR luminosity is estimated from a multicomponent (AGN and starburst) fit to the FIR using {\tt Decompir} software \citep{mullaney2011} and following the method described by \citep{seymour2012}. We use the 24 $\mu$m flux density upper limit to constrain the short wavelength part of the IR SED and find the AGN contribution to the total FIR luminosity is  $\lesssim$10\%. ALESS65.1 therefore appears to be dominated by star-formation processes and the AGN contribution to [CII] and  $L_{\rm FIR}$ is minimal. 

\begin{table}
\renewcommand{\thefootnote}{\alph{footnote}}
\begin{minipage}{\columnwidth}
\centering
\caption{Observed and derived properties of ALESS65.1}
\begin{tabular}{lcc}  \hline
Parameter & Value & Reference \\ \hline
$z_{\rm [CII]}$ & 4.4445 $\pm$ 0.0005 & S12\\
$I_{\rm [CII]}$ & 5.4 $\pm$ 0.7 Jy km s$^{-1}$ & S12\\  
FWHM$_{\rm [CII]}$ & 470 $\pm$ 35 km s$^{-1}$ & S12\\
$L_{\rm [CII]}$ & (3.2 $\pm$ 0.4) $\times$ $10^9$ ${\rm L}_\odot$ & S12 \\
$L_{\rm FIR}$\footnotemark[1] & (1.38 $\pm$ 0.28) $\times$ $10^{12}$ ${\rm L}_\odot$ & S12 \\
$I_{\rm CO(2-1)}$ & $< 0.11$ Jy km s$^{-1}$ (3$\sigma$) & this paper \\  
$M_{\rm gas}$ & $<1.7 \times 10^{10}$ ${\rm M}_\odot$  (3$\sigma$) & this paper \\
$L_{\rm CO(2-1)}$ &  $<$ 8.5 $\times$ $10^{6}$ ${\rm L}_\odot$ (3$\sigma$) & this paper \\
$L'_{\rm CO(2-1)}$ &  $<$ 2.2 $\times$ $10^{10}$ K km s$^{-1}$ pc$^2$ & this paper \\
\hline
\end{tabular}
\footnotetext{$^{\rm a}$ converted from $L_{\rm IR}$(8 --- 1000$\mu$m) using $L_{\rm FIR}$(42 -- 122 $\mu$m) = $L_{\rm IR}$/1.45 (Stacey et al. 2010; De Breuck et al. 2011)}
\end{minipage}
\end{table}

In examining the PDR physical conditions, we multiply the $^{12}$CO(2--1) flux by a factor of two to account for the line being optically thick, and hence only the $^{12}$CO emission coming from the illuminated PDR side is seen \citep{kaufman1999, hailey-dunsheath2010}. We also multiply the [CII] flux by a factor of 0.7 to remove non-PDR contributions (e.g. \citealp{hailey-dunsheath2010, stacey2010}). The $^{12}$CO geometry correction applies to all galaxies in Figure 2, and so does not affect the relative position of ALESS65.1 on the diagram compared to other galaxies. 
Using the \cite{kaufman1999} models,  we find ALESS65.1 has $G_0 \sim 10^3$ and $n \lesssim10^5$ cm$^{-3}$ (Figure 2). Such a FUV radiation field is on the high end of those seen in low redshift normal galaxies, but it is consistent with the strong FUV fields seen in local starbursts,  nearby ULIRGs, and some $z > 1$ galaxies.  This limit on $G_0$ and $n$ implies a PDR temperature $\gtrsim$300 K \citep{kaufman1999}. Using Equation 1 from \cite{hailey-dunsheath2010}, we estimate the atomic gas associated with the PDR to be greater than 3 $\times$ $10^9$ $M_{\odot}$. We note that this atomic mass estimate is very uncertain, and it is highly dependent on $n$. The atomic gas mass $M_a \sim 2.9 \times 10^{9} + (3.2 \times 10^{12}/n)$, so for $n = 10^3$ cm$^{-3}$ the atomic mass would be $\sim$6 $\times$ $10^9$ $M_{\odot}$.

\begin{figure*}
\includegraphics[width=0.75\textwidth]{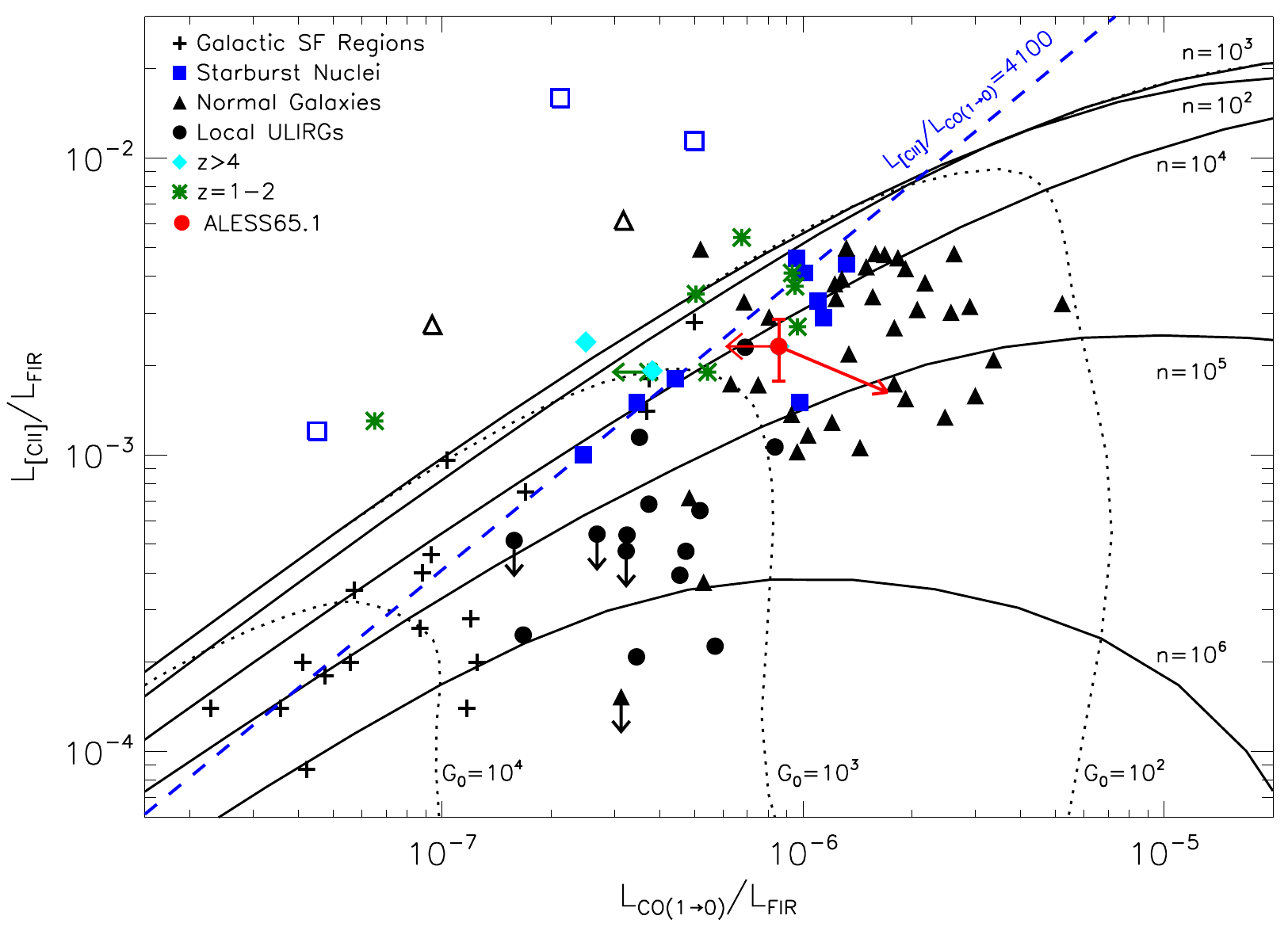}
\caption{$L_{\rm [CII]}/L_{\rm FIR}$  versus $L_{\rm CO(1-0)}/L_{\rm FIR}$ for ALESS65.1 (red point with upper limit on $L_{\rm CO(1-0)}/L_{\rm FIR}$) compared with Galactic star forming regions, starburst nuclei, normal galaxies, local ULIRGs, and high redshift ($z > 1$) sources. Empty symbols indicate low metallicity sources, which lie at high $L_{\rm [CII]}/L_{\rm CO(1-0)}$. Black lines represent the solar metallicity PDR model calculations for gas density ($n$) and FUV field strength ($G_0$) from Kaufman et al. (1999). This figure is adapted from Stacey et al. (2010) with additional data from De Breuck et al. (2011) and Walter et al. (2012). ALESS65.1 has a higher $L_{\rm [CII]}/L_{\rm FIR}$ ratio than that found in local ULIRGs, similar to the other two $z > 4$ SMGs, but its $L_{\rm [CII]}/L_{\rm CO}$ ratio is similar to local starbursts and other $z > 1$ sources. }
\label{fig:cii_co}
\end{figure*}

 ALESS65.1 and the other two $z > 4$ SMGs shown in Figure 2, LESS J033229.4 (De Breuck et al. 2011) and HDF850.1 (Walter et al. 2012), have a similar FUV radiation field, $G_0$, to local starbursts, but have much higher FIR luminosity, leading to suggestions that they are scaled-up versions of local starbursts. For a given $L_{\rm FIR}$ the size of the emission region will increase for smaller $G_0$. Following Stacey et al. (2010), we scale up from M82 using two laws from \cite{wolfire1990} to constrain the size: $G_0 \propto \lambda L_{\rm FIR}/D^3$ if the mean free path of a UV photon $\lambda$ is small and  $G_0 \propto L_{\rm FIR}/D^2$ if the mean free path of a UV photon is large. Applying these relations and using $G_0 = 10^3$ for ALESS65.1 yields a diameter of 1.1 -- 2.1 kpc. This is consistent with the marginally resolved [CII] data which shows ALESS65.1 has a possible extent of 3.3 $\pm$ 1.7 kpc (S12). The same scaling relation applied to HDF850.1 results in a diameter of 1.8 -- 4.6 kpc, which is consistent with the observed 5.7 $\pm$ 1.9  kpc extent of the [CII] emission region (Walter et al. 2012). Similarly, LESS J033229.4 has an extent of $\sim$ 4 kpc (Coppin et al. 2010; De Breuck et al. 2011). Thus the starburst in all three $z > 4$ SMGs appears to be extended over galactic scales.  
  
Local starbursts and Galactic OB star forming regions lie on a line with [CII]/CO luminosity ratios of about 4100 in Figure 2. Higher [CII]/CO ratios can also be found in low metallicity systems, such as 30 Doradus in the LMC, where the size of the [CII] emitting envelope of the cloud (relative to the the CO emitting core) is much larger than in more metal-rich systems \citep{stacey1991}. Metallicity may also affect the [CII]/CO ratios of the highest redshift galaxies. For example, LESS J033229.4 at $z = 4.76$ has a very high [CII]/CO ratio of $\approx$ $10^4$ and was initially thought to have sub-solar metallicity \citep{debreuck2011}, but more recent [NII] observations suggest that the metallicity of this galaxy is solar \citep{nagao2012}. 
ALESS65.1 has  $L_{\rm [CII]}/L_{\rm CO} \gtrsim 2700$, which is consistent with low metallicity, but it has an extent of several kpc so therefore the enhanced [CII] emission is more likely to be due to the extended star forming region.  An extended (or less-dense) interstellar medium would have an increased fraction of UV photons available to ionize and excite gas, increasing the relative intensity of fine structure lines (e.g. \citealp{gracia-carpio2011}). 

We have assumed a single-phase ISM in this work. Multi-phase ISMs are commonly required to explain the $^{12}$CO line ratios observed in local galaxies (e.g. \citealp{guesten1993, ward2003}). In some cases a single-component ISM is unable to explain the $^{12}$CO-excitation ladder (or spectral line energy distribution) observed in high redshift SMGs \citep{carilli2010,harris2010, danielson2011}. 
These studies found that the ISM in such sources is best described by a warm compact component surrounded by a cooler more extended one. 
Support for such a multi-component geometry in high-redshift sources has recently been found in the spatially-resolved $^{12}$CO study of  the gravitationally-magnified submm galaxy SMM\,J2135$-$0102 \citep{swinbank2011,danielson2011}.  Future higher-resolution continuum and line studies with ALMA may uncover similar evidence in ALESS\,65.1, but the current data provide no direct evidence. 

\section{Conclusion}

We have observed ALESS65.1 for 20 hours to search for $^{12}$CO(2--1) emission in this $z = 4.44$ submillimetre galaxy. We detect no $^{12}$CO(2--1) emission in a spectrum reaching a rms sensitivity of 0.08 mJy beam$^{-1}$ per 200 km s$^{-1}$ channel. 

Adopting the FWHM from the ALMA detection of [CII] in ALESS65.1, we estimate a $3 \sigma$ limit to the $^{12}$CO(2--1) luminosity of $L_{\rm CO(2-1)} <8.5 \times 10^{6}$ ${\rm L}_\odot$  and a cold gas mass upper limit of $M_{\rm gas} <1.7 \times 10^{10}$ ${\rm M}_\odot$.  This implies a gas depletion timescale in ALESS65.1 of $< 50$ Myr, comparable to other $z > 4$ SMGs and consistent with this high redshift population being the progenitors of $z > 2$ red-and-dead galaxies.

We examine the physical conditions of the gas in ALESS65.1 using the $L_{\rm [CII]}/L_{\rm FIR}$  versus $L_{\rm CO(1-0)}/L_{\rm FIR}$ diagram. We find ALESS65.1 has a strong FUV field comparable to local starbursts. The observed [CII] to FIR ratio, $L_{\rm [CII]}/L_{\rm FIR} = 2.3 \times 10^{-3}$, is high compared to local ULIRGs (as noted by S12). Combined with $L_{\rm [CII]}/L_{\rm CO} \gtrsim 2700$, this high [CII] to FIR ratio is consistent with ALESS65.1 having more extended regions of intense star-formation than local ULIRGs. A possible, but less likely scenario, is ALESS65.1 has low metallicity gas. 

Measurements of [CII] and $^{12}$CO of a larger sample are needed to confirm whether $z > 4$ starbursts have enhanced [CII] emission compared to local galaxies, and whether this is because of metallicity effects, the relative size of PDR regions, a combination of the two, or other effects.  Future surveys by ALMA will shed further light on the physical conditions of the gas in star forming galaxies in the early universe.  

\section*{Acknowledgments}

AK and IRS acknowledge support from the STFC and IRS also acknowledges support from the Leverhulme Trust. TRG acknowledges support from an STFC Advanced Fellowship. 
KC acknowledges support from the endowment of the Lorne Trottier Chair in Astrophysics and Cosmology at McGill and the Natural Science and Engineering Research Council of Canada (NSERC). NS is the recipient of an Australian Research Council Future Fellowship. KK thanks the Swedish Research Council for support.
The Australia Telescope is funded by the Commonwealth of Australia for operation as a National Facility managed by CSIRO.
This paper makes use of ALMA  data from the project ADS/  JAO.ALMA\#2011.1.00294.S. ALMA is a partnership of ESO (representing   its member states),  NSF (USA) and NINS (Japan), together with NRC   (Canada) and NSC and ASIAA (Taiwan), in cooperation  with the Republic of Chile. The Joint ALMA Observatory is operated by ESO, AUI/NRAO and NAOJ.

\bibliographystyle{mn2e}
\bibliography{refs}

\bsp

\label{lastpage}

\end{document}